\documentclass[reprint, superscriptaddress, secnumarabic, amssymb, nobibnotes, aps, prl]{revtex4-1}

\usepackage{graphicx}
\usepackage{epstopdf}
\usepackage[T1]{fontenc}
\usepackage[latin9]{inputenc}
\usepackage{amsbsy}
\usepackage{gensymb}
\usepackage[T1]{fontenc}
\usepackage[latin9]{inputenc}
\usepackage{amsmath}
\usepackage{amssymb}
\usepackage{bbm}
\usepackage{braket}
\usepackage{xcolor}
\allowdisplaybreaks
\usepackage{graphicx}
\usepackage[colorlinks=true]{hyperref}  
\hypersetup{
    bookmarks=true,         
    unicode=false,          
    pdftoolbar=true,        
    pdfmenubar=true,        
    pdffitwindow=false,     
    pdfstartview={FitH},    
    pdftitle={Superconductivity with high upper critical field in Ta-Hf Alloys},    
    pdfauthor={R. P. Singh},     
    pdfsubject={},   
    pdfcreator={},   
    pdfproducer={}, 
    pdfkeywords={} {} {}, 
    pdfnewwindow=true,      
    colorlinks=true,       
    linkcolor=blue, 
    citecolor=blue,        
    filecolor=magenta,      
    urlcolor=blue           
    }
\usepackage[normalem]{ulem}

\newcommand{\equref}[1]{Eq.~(\ref{#1})}

\newcommand{\figref}[1]{Fig.~\ref{#1}}

\newcommand{\tableref}[1]{Table~\ref{#1}}

\renewcommand{\approx}{\simeq}

\begin{document}

\title{\textrm{Superconductivity with high upper critical field in Ta-Hf Alloys}}

\author{P. K. Meena}
\affiliation{Department of Physics, Indian Institute of Science Education and Research Bhopal, Bhopal, 462066, India}

\author{S. Jangid}
\affiliation{Department of Physics, Indian Institute of Science Education and Research Bhopal, Bhopal, 462066, India}

\author{R. K. Kushwaha}
\affiliation{Department of Physics, Indian Institute of Science Education and Research Bhopal, Bhopal, 462066, India}

\author{R. P. Singh}
\email[]{rpsingh@iiserb.ac.in}
\affiliation{Department of Physics, Indian Institute of Science Education and Research Bhopal, Bhopal, 462066, India}

\begin{abstract}
High upper-critical field superconducting alloys are required for superconducting device applications. In this study, we extensively characterized the structure and superconducting properties of alloys Ta$_{x}$ Hf$_{1-x}$ (x = 0.2, 0.4, 0.5, 0.6 and 0.8). The substitution of Hf (T$_{C}$ = 0.12 K, type-I superconductor) with Ta (T$_{C}$ = 4.4 K, type-I superconductor) shows an anomalous enhancement of T$_{C}$ with variation of composition. Interestingly, all compositions exhibited strongly coupled bulk type-II superconductivity with a high upper critical field. In particular, for compositions x = 0.2, and 0.4, the upper critical field (H$_{C2}$) approached the Pauli limiting field.
\end{abstract}
\maketitle
\section{INTRODUCTION}
Superconductivity, a quantum phenomenon with significant practical applications, has recently led to the exploration of unconventional superconductors that exhibit remarkable properties that differ from the conventional Bardeen-Cooper-Schrieffer (BCS) model \cite{BCS}. These unconventional superconductors exhibit remarkable features, such as an upper critical field comparable to or exceeding the Pauli paramagnetic field, strong electron-phonon interactions, the presence of gap nodes, and the breaking of time-reversal symmetry (TRSB) \cite{Unconventional-SC1, Maki}. The strength of spin-orbit coupling (SOC) in the materials under investigation plays a pivotal role in the emergence of unconventional superconductivity \cite{USC-Pairings1, USC-Pairings2, TI, DW, TIS, TS1, SOC1}. Superconductors based on heavier elements with higher atomic numbers, particularly those in the 5d series, tend to exhibit robust spin-orbit coupling (SOC $\propto$ Z$^{4}$), giving rise to these unconventional superconducting behaviors \cite{IrGe, Zr2Ir, TaC, TaOsSi, PbTaSe2, LiPtB1, LiPtB2}.\\

The Ta-Hf binary alloy is a notable example of a 5d superconducting alloy that combines two type-I superconductors, Ta and Hf. It is anticipated to exhibit strong spin-orbit coupling due to the high atomic numbers of its constituent elements. Surprisingly, this alloy displays remarkable behavior in the upper critical field. By partially substituting Hf for Ta, it is possible to modify the strength of spin-orbit coupling \cite{Ti-Nb-Ta}. When Hf atoms are introduced into Ta, which has a T$_{C}$ of 4.2 K, the superconducting transition temperature increases to 6.7 K in the Hf-Ta alloy with approximately 40 \% Hf \cite{Hf}. The relationship between the density of states at the Fermi level, the electron-phonon coupling, and the number of valence electrons (d shell) per atom has been correlated with the superconducting transition temperature (T$_C$) \cite{d-shellmetals1, d-shellmetals2}. However, T$_C$ exhibits non-monotonic behavior with valence electron counts, a trend observed in other binary alloy superconductors \cite{Binaryalloy1, Mo-Re1, Mo-Re2, Nb-Ta, HT2, Nb-Zr, d-shellmetals1, Nb-Mo}. These metallic alloys possess both metallic properties and a high upper critical field, making them highly promising for practical superconducting devices.

Despite studies on Ta-Hf binary alloys \cite{HC2, HT1}, the mechanisms responsible for the enhanced critical temperature (T$_C$) and the high upper critical field behavior in these alloys have not been fully understood, primarily due to incomplete characterization. Unravelling these mechanisms could provide valuable insight and enable the synthesis of metallic alloys with enhanced properties suitable for practical applications. Thus, the Ta-Hf alloy is a promising candidate for in-depth investigations into its superconducting properties, offering a platform to better comprehend other binary superconducting behaviours.

\begin{figure*}
\includegraphics[width=2.0\columnwidth]{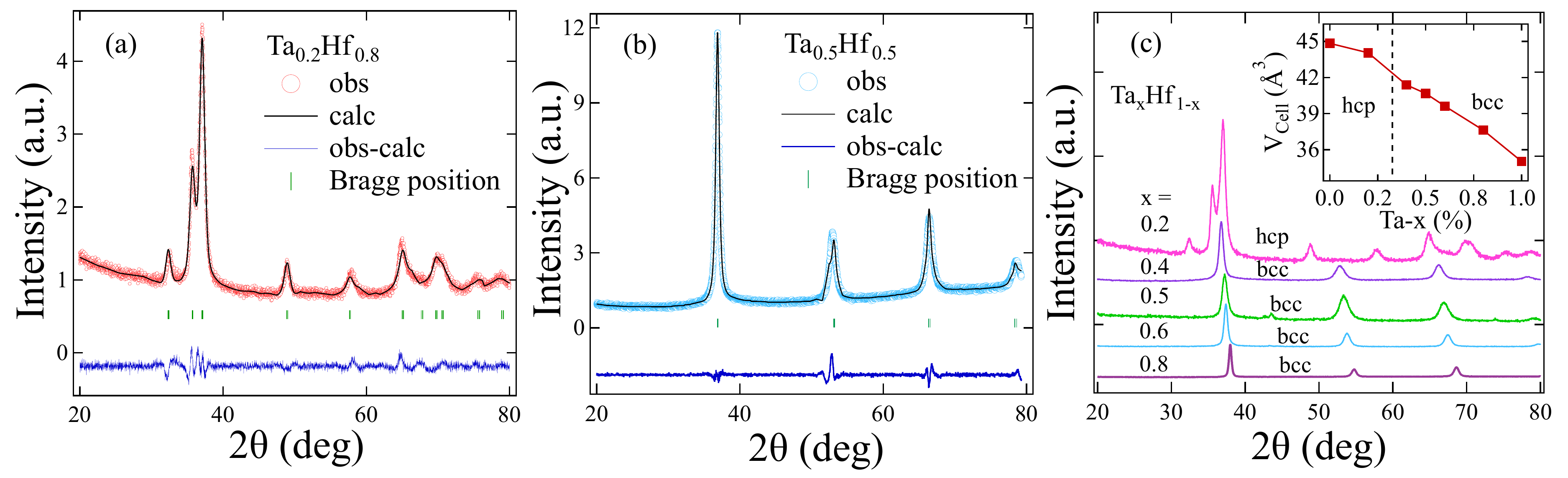}
\caption {\label{XRD pattern} Powder XRD patterns with refinement for binary alloys, (a) Ta$_{0.2}$Hf$_{0.8}$. (b) Ta$_{0.5}$Hf$_{0.5}$. (c) All compositions XRD patterns for Ta$_{x}$Hf$_{1-x}$. The inset shows the refined cell volume as a function of the Ta content $x$.}
\label{XRD pattern.jpg}    
\end{figure*}

In this study, we investigate the superconducting properties of binary Ta$_{x}$Hf$_{1-x}$ alloys, where $x$ takes values of 0.2, 0.4, 0.5, 0.6, and 0.8. Our analysis included electrical resistivity, DC magnetization, and heat capacity measurements, allowing us to construct the phase diagram for these alloys. Throughout the entire range of the solid solution, we observed the coexistence of two crystal structures: W-$bcc$ and Mg-$hcp$. Importantly, all compositions of Ta$_{x}$Hf$_{1-x}$ exhibited bulk type-II superconductivity. Notably, the presence of strongly coupled superconductivity and an upper critical field comparable to the Pauli limiting field suggests the potential occurrence of unconventional superconductivity.

\section{EXPERIMENTAL DETAILS}
Polycrystalline samples of Ta$_{x}$Hf$_{1-x}$ were prepared by arc melting using high-purity Ta(99.99\%) and Hf(99.99\%) metals in stoichiometric ratios in a high-purity Ar(99.99\%) environment on a water-cooled copper hearth. Ingots were repeatedly remelted and flipped to enhance chemical homogeneity with minimal mass loss. A titanium button was used as a getter to remove residual oxygen in the chamber. Crystal structure and phase purity were verified using powder X-ray diffraction (PXRD) on a PANalytical diffractometer equipped with Cu$K_{\alpha}$ radiation ($\lambda$ = 1.54056 $\text{\AA}$). Magnetization measurements were performed using the Magnetic Property Measurement System (MPMS3; Quantum Design). Specific heat and electrical resistivity measurements were carried out using the Physical Property Measurement System (PPMS; Quantum Design).

\section{EXPERIMENTAL RESULTS}
\subsection{Sample characterization}
All synthesized Ta$_{x}$Hf$_{1-x}$ alloys (with $x$ = 0.2, 0.4, 0.5, 0.6, and 0.8) are found to be in a pure phase and crystallize into two distinct crystal structures. \figref{XRD pattern}(a,b) shows representative XRD patterns for polycrystalline binary alloys with $x$ = 0.2 and 0.5, while the XRD patterns of other alloys closely resemble those of $x$ = 0.5. We used FullProf Rietveld software \cite{FullProf} to analyze the XRD patterns, revealing that the samples can be well indexed by the Mg-$hcp$ structure with space group P6$_{3}$/$mmc$ for $x$ = 0.2 and the W-$bcc$ structure with space group I$m\bar{3}m$ for the remaining compositions. \figref{XRD pattern}(c) displays the XRD patterns for all compounds ($x$ = 0.2, 0.4, 0.5, 0.6, and 0.8), while the refined structural lattice parameters for these compounds are summarized in \tableref{tbl: SC parameters}. The lattice constants obtained for some binary compounds are consistent with previous studies in the literature \cite{HfTa-phasediagram}, whereas others are reported for the first time in this work. The inset of \figref{XRD pattern}(c) shows a linear decrease in $V_{cell}$ with increasing $x$, which can be attributed to the smaller atomic radius of Ta compared to Hf.

\begin{figure*} 
\includegraphics[width=2.0\columnwidth]{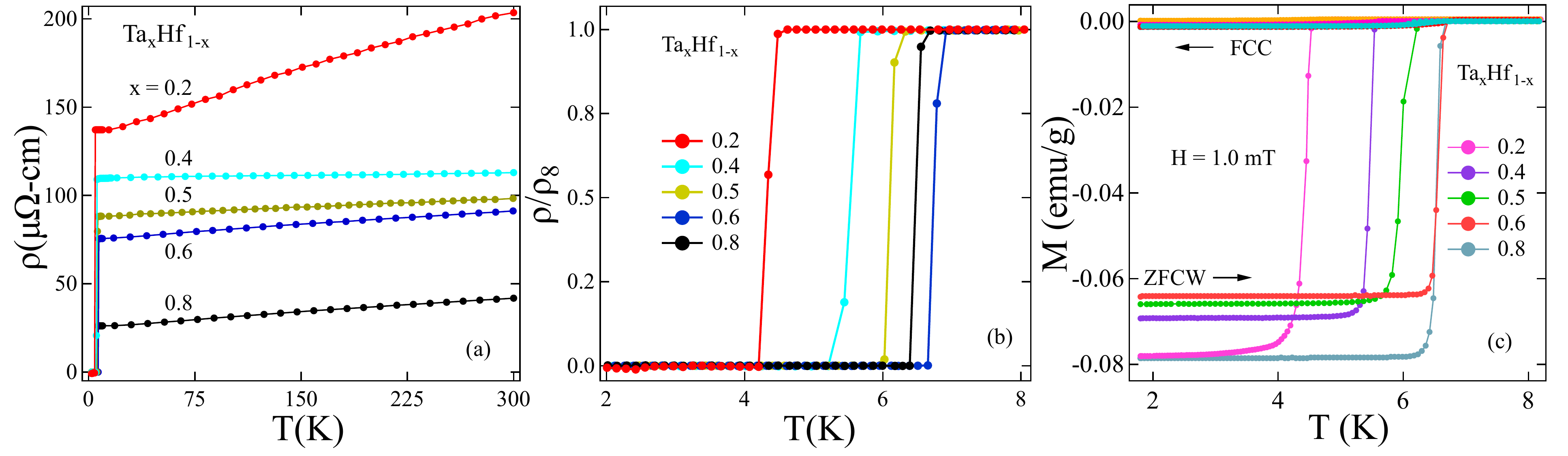}
\caption {\label{RT and MT} (a) Temperature variation of electrical resistivity $\rho(T)$ at H = 0 mT for Ta$_{x}$Hf$_{1-x}$ ($x$ = 0.2, 0.4, 0.5, 0.6, and 0.8). (b) A zoomed-in view of the normalized resistivity ($\rho$/$\rho_{8}$) highlights the superconducting drop observed in all compositions. (c) Magnetization data obtained during field-cooled cooling (FCC) and zero field-cooled heating (ZFCW) measurements in an applied field of H = 1.0 mT show the presence of superconductivity below $T_{C}$ for all binary Ta$_{x}$Hf$_{1-x}$ compositions.} 
\end{figure*} 

\subsection{ Superconducting and normal state properties}
Superconductivity in the Ta$_{x}$Hf$_{1-x}$ alloys (where $x$ = 0.2, 0.4, 0.5, 0.6, and 0.8) was confirmed by measurements of electrical resistivity and DC magnetization. Temperature-dependent electrical resistivity ($\rho(T)$) was measured from 300 K to 1.9 K at zero magnetic field (H = 0 mT), as shown in \figref{RT and MT}(a). The resistivity exhibited a slight temperature variation above the superconducting transition temperature (T$_{C}$), indicating the weak metallic character of the Ta-Hf alloys \cite{Nb4Rh2C, LiPdCuB}. The values of the residual resistivity ratio (RRR), defined as the ratio of resistivity at 300 K to that at 8 K, were found to be relatively small (RRR $\approx$ 1-2) for all compositions, suggesting the presence of disorder in the binary alloy. The RRR values follow a similar pattern observed in other binary alloys \cite{NbTi-thinfilm} and are provided in \tableref{tbl: SC parameters}. \figref{RT and MT}(b) presents an expanded plot of the normalized electrical resistivity data at zero field, clearly showing the superconducting transitions corresponding to different compositions of the Ta-Hf alloy. The superconducting transition temperature (T$_{C}$) varies non-linearly with the Hf (or Ta) concentration, ranging from 0.12 K for pure Hf to 6.7 K for a Ta concentration of 60\% in solid solution, with a T$_{C}$ of 4.2 K for pure Ta.

The magnetic moment variations with temperature were measured at an applied field of 1.0 mT using two different modes: zero-field cooled warming (ZFCW) and field-cooled cooling (FCC). Magnetization data for all samples exhibited the emergence of diamagnetic behavior at different transition temperatures (T$_{C}$), consistent with resistivity measurements, as shown in \figref{RT and MT}(c). The Ta-Hf binary alloy displayed a distinct dome-shaped behavior, similar to that observed in the Ti-V \cite{Ti-V} and Zr-Nb \cite{Zr-Nb} binary alloys. The maximum and minimum T$_{C}$ values of 6.7 K and 4.5 K were observed for compositions corresponding to 60\% and 20\% Ta concentration, respectively. The separation between the ZFCW and FCC modes in the magnetization data below T$_{C}$ indicates a strong magnetic flux pinning. The respective T$_{C}$ values obtained from DC magnetization consistent with the electrical transport measurements and values are summarized in \tableref{tbl: SC parameters}.
 
\begin{figure*}
\includegraphics[width=2.0\columnwidth]{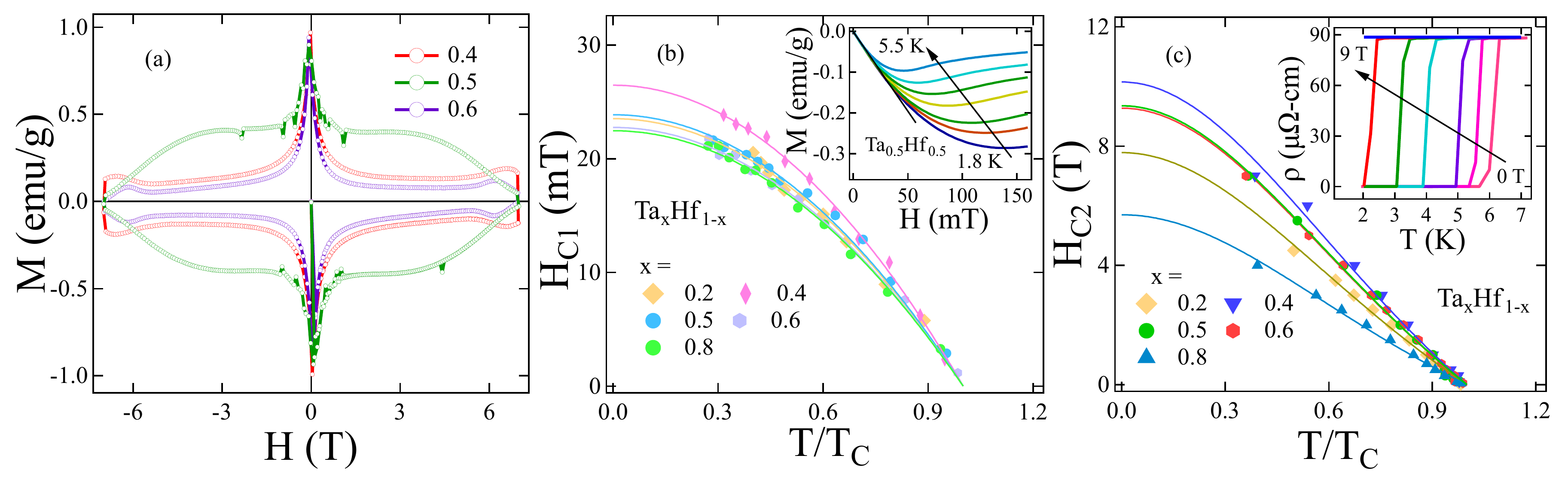}
\caption {\label{Fig3} (a) The magnetization of Ta$_{x}$Hf$_{1-x}$ alloys was measured in the magnetic field range of -7 T to 7 T. The fishtail effect near a magnetic field of 6 T is evident in the data. (b) The temperature dependence of the lower critical field (H$_{C1}$) was determined from magnetization measurements. The H$_{C1}$ values were fitted using the Ginzburg-Landau (GL) equation for Ta$_{x}$Hf$_{1-x}$ alloys. The inset shows the magnetization curve (M(H)) for Ta$_{0.5}$Hf$_{0.5}$ at various temperatures. (c) The temperature dependence of the upper critical field (H$_{C2}$) was obtained from resistivity measurements. The H$_{C2}$ values were fitted using the GL equation. The inset shows the temperature-dependent resistivity ($\rho(T)$) for Ta$_{0.5}$Hf$_{0.5}$ at different magnetic fields.}
\end{figure*}

Magnetization versus field (M-H) measurements were conducted for Ta$_{x}$Hf$_{1-x}$ alloys to confirm their type-II behavior. \figref{Fig3}(a) displays the M-H curves for $x$ = 0.4, 0.5, and 0.6, revealing the presence of the fishtail effect. The composition $x$ = 0.5 also exhibits a flux jump in the magnetization loop. These unconventional vortex states, typically observed in high-T$_{C}$ oxides and certain two-dimensional superconducting materials \cite{Fishtail-effect1, Fishtail-effect2, Fishtail-effect3, Fishtail-effect4}, suggest the influence of strong disorder in the material.

The lower critical field, H$_{C1}$(0), was estimated by measuring the low field M-H. The temperature dependence of H$_{C1}$ is determined by identifying the point at which the M(H) curves deviate from the Meissner line, as shown in the inset of \figref{Fig3}(b) for Ta$_{0.5}$Hf$_{0.5}$. The temperature variation of the H$_{C1}$ values for all compositions is presented in \figref{Fig3}(b). Utilizing the Ginzburg-Landau (GL) theory of phase transition, H$_{C1}$(0) values for Ta$_{x}$Hf$_{1-x}$ can be obtained by fitting the expression for H$_{C1}$(T),
\begin{equation}
H_{C1}(T)=H_{C1}(0)\left[1-\left(\frac{T}{T_{C}}\right)^{2}\right].
\label{eqn1:HC1}
\end{equation}

The estimated H$_{C1}$(0) is in the range of 20-30 mT for all compositions of Ta$_{x}$Hf$_{1-x}$ ($x$ = 0.2, 0.4, 0.5, 0.6, and 0.8). Temperature-dependent magnetization and resistivity data were also taken in various applied external magnetic fields up to 7.0 T and 9.0 T, respectively, to determine the upper critical field H$_{C2}$(0). The observed changes in T$_{C}$ with the increasing applied magnetic field, as shown in the inset of \figref{Fig3}(c) for Ta$_{0.5}$Hf$_{0.5}$, are interpreted as the upper critical field H$_{C2}$(T). Temperature-dependent H$_{C2}$ data were fitted using Ginzburg-Landau (GL) expression for respective compositions to determine the H$_{C2}$(0) values as given in \equref{eqn2:HC2} and the fitting is shown by solid lines in \figref{Fig3}(c). 

\begin{equation}
H_{C2}(T) = H_{C2}(0)\left[\frac{(1-t^{2})}{(1+t^{2})}\right],  
   \quad  \text{where} \;  t = \frac{T}{T_{C}}
\label{eqn2:HC2}
\end{equation}
The estimated H$_{C2}$(0) values through aforementioned measurements are given in the \tableref{tbl: SC parameters} for all the composite alloys. Maximum H$_{C2}$(0) values are obtained as 10.43 T for Ta$_{0.4}$Hf$_{0.6}$ and exhibit a steady decrease with increasing Ta content in replacement of Hf content \cite{HC2}. 

In type-II superconductors, the presence of a magnetic field leads to the destruction of superconductivity due to two main effects: the orbital limiting field and the Pauli paramagnetic field. In situations where both effects are significant, the temperature dependence of the upper critical field can be described by the Werthamer-Helfand-Hohenberg (WHH) theory. This theory considers the combined influences of spin paramagnetism and spin-orbit interaction. In the absence of Pauli spin paramagnetism and the spin-orbit interaction, the WHH theory provides the following equation \cite{WHHM1, WHHM2}:
\begin{equation}
H^{orb}_{C2}(0) = -\alpha T_{C} \left.{\frac{dH_{C2}(T)}{dT}}\right|_{T=T_{C}} 
\label{eqn3:WHH}
\end{equation}
which gives the orbital limit field of the Cooper pair. The constant $\alpha$, is the purity factor of 0.693(0.73) for dirty(clean) limit superconductors. For the dirty limit condition, we obtained the H$^{orb}_{C2}$(0) values, which are summarized in \tableref{tbl: SC parameters}. But the latter $\lambda_{SO}$ has been shown to increase with increasing atomic numbers of the composing elements \cite{HT1} and is thus expected to be high for Ta$_{x}$Hf$_{1-x}$, so the spin-orbit scattering effect cannot be ignored. However, the Pauli limit of the upper critical field \cite{Pauli1,Pauli2}, can be calculated by following relation H$_{C2}^p$(0) = 1.84 T$_{C}$ within the BCS theory. We calculated the H$_{C2}^p$(0) values using the estimated T$_{C}$ of the resistivity curves as 8.28, 10.34, 11.59, 12.32, 12.14 T for Ta$_{x}$Hf$_{1-x}$ ($x$ = 0.2, 0.4, 0.5, 0.6, and 0.8). Interestingly, the value of H$_{C2}$(0) is comparable to Pauli's limiting field for $x$ = 0.2, 0.4, indicating the unconventionality present in these compounds. Similar results have been observed in certain superconductors, including Chevrel phase \cite{Cheveral}, A15 compounds \cite{Ti3Sb}, and various Re-based noncentrosymmetric superconductors \cite{Re5.5Ta, NbReSi1, NbReSi2}. To further quantify the impact of spin paramagnetic effects, we calculated the Maki parameter ($\alpha_{m}$), which is given in the following expression:
\begin{equation}
\alpha = \sqrt{2} \frac{H^{orb}_{C2}(0)}{H_{C2}^p(0)} 
\label{eqn4:Maki}
\end{equation}

The calculated values of $\alpha_{m}$ are provided in \tableref{tbl: SC parameters}. The observed variation of $\alpha_{m}$ with doping can be attributed to the interplay between spin-orbit coupling and doping-induced disorder, arising from differences in atomic numbers (Z) and atomic radii of the constituent elements in the Ta-Hf alloy. Two length scales of characteristics of a superconductor were determined: penetration depth $\lambda_{GL}$(0) and Ginzburg-Landau coherence length $\xi_{GL}$(0) from the given relations using the value of H$_{C1}$(0) and H$_{C2}$(0), 

\begin{equation}
H_{C2}(0) = \frac{\Phi_{0}}{2\pi\xi_{GL}^{2}(0)},
\label{eqn5:coherence}
\end{equation}
\begin{equation}
H_{C1}(0) = \frac{\Phi_{0}}{4\pi\lambda_{GL}^2(0)}\left( ln \frac{\lambda_{GL}(0)}{\xi_{GL}(0)} + 0.12\right);  
\label{eqn6:lamda}
\end{equation}
where $\Phi_{0}$ denotes the fluxoid quantum ($\Phi_{0}$ = 2.07 $\times$10$^{-15}$ T m$^{2}$) \cite{tin}. Subsequently, the GL parameter is defined as $k_{GL}$ = $\frac{\lambda_{GL}(0)}{\xi_{GL}(0)}$, which signifies the type of superconductivity (either I or II), was also calculated for each composition. Moreover, the relation $H_{C1}(0) H_{C2}(0)$ = $H_{C}^2 (0)$ $ ln(k_{GL}(0)+0.08)$ can be used to compute thermodynamic critical field H$_{C}$ at 0 K using the value of H$_{C1}$(0), H$_{C2}$(0) and $k_{GL}$. The obtained values of H$_{C}$ are 190-300 mT for all compositions. All physical parameters $\xi_{GL}$(0), $\lambda_{GL}$(0), and $k_{GL}$ are summarized in \tableref{tbl: SC parameters}. The estimated values of $\xi_{GL}$(0) and $\lambda_{GL}$(0) are in the range of 5-8 nm and 140-160 nm, respectively, same as Re-doped MoTe$_{2}$ \cite{ReMoTe2}, Ru and Ir doped LaRu$_{3}$Si$_{2}$ \cite{LaRu3Si2}. The large values of $k_{GL}$ in the range of 23-27 \cite{V-Ti} for binary compounds, suggests that Ta$_{x}$Hf$_{1-x}$ have strong type-II superconductivity.

\begin{figure} [b]
\includegraphics[width=1.0\columnwidth]{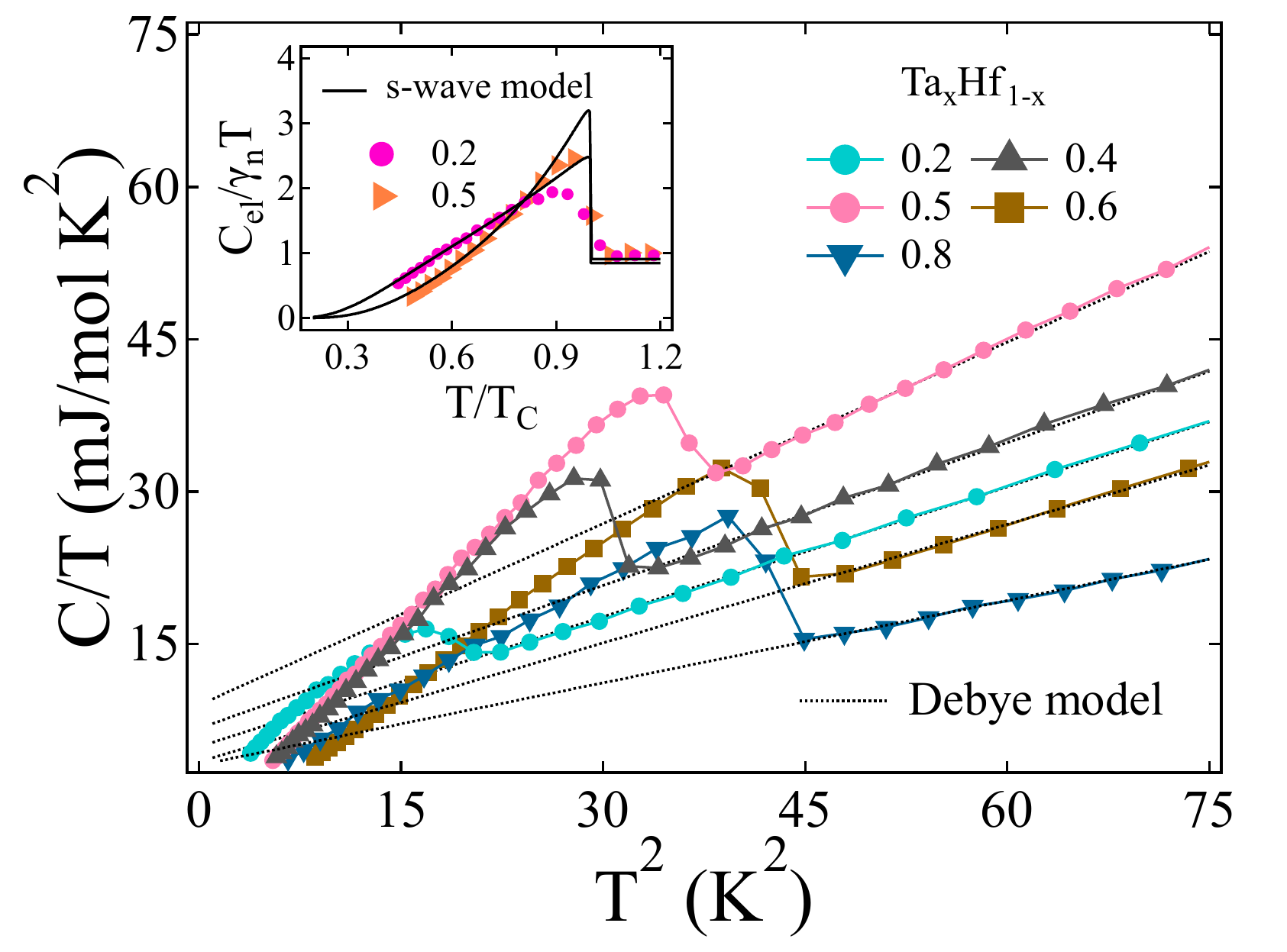}
\caption {\label{Fig4.jpg}
C/T versus $T^2$ measured at H = 0 mT and fitted in normal state using relation C = $\gamma_{n}$T + $\beta$ T$^{3}$. Inset shows normalized electronic specific heat, which is well described with the isotropic BCS model.}
\end{figure}

Heat capacity measurements were conducted for all compositions to further characterize superconductivity in Ta$_{x}$Hf$_{1-x}$ alloys. As depicted in Figure \ref{Fig4.jpg}, the heat capacity plots exhibit a discontinuity that signifies the transition from normal to superconducting state. The determined T$_{C}$ values of the specific heat are consistent with all the resistivity and magnetization measurements of the compounds. Low-temperature specific heat can be described using the Debye relation, C = $\gamma_{n}$T + $\beta$T$^{3}$, where the term $\gamma_{n}$T represents the electronic contribution, and the term $\beta$T$^{3}$ corresponds to the phononic contribution. By extrapolating the behavior of the normal state to low temperatures, the Sommerfeld coefficient ($\gamma_{n}$) and the Debye constant ($\beta$) can be estimated.

\begin{table*}
\scriptsize
\caption{ Ta$_{x}$Hf$_{1-x}$ ($x$ = 0.2 - 0.8) Space group, refined lattice parameters, and cell volume derived from X-ray refinement as well as superconducting and normal-state parameters derived from magnetization, resistivity, and specific heat measurements.}
\label{tbl: SC parameters}
\setlength{\tabcolsep}{1pt}
\begin{center}
\begin{tabular}{p{0.13\linewidth}p{0.13\linewidth}p{0.12\linewidth}p{0.12\linewidth}p{0.12\linewidth}p{0.12\linewidth}p{0.10\linewidth}}
\hline
Parameter &unit & 0.2 & 0.4 & 0.5 &0.6 & 0.8\\
 \hline
Space group & &P6$_3$/$mmc$ &&I$m\bar{3}m$ \\
a = b  &\AA&$\text{3.1849}$&3.4586&3.4392&3.4034&3.3509\\
c &\AA &$\text{5.0143}$&3.4586&3.4392&3.4034&3.3509\\
V$_{cell}$ &\AA$^{3}$&44.05&41.37&40.67&39.62&37.62\\
T$_{C}$ & K &4.5& 5.6&6.1&6.7&6.6\\
H$_{C1}(0)$ & mT & 23.52 & 26.40 & 23.87 & 22.6 & 22.51\\ 
H$_{C2}^{res}$(0) & T &7.72 & 10.43&9.3&9.27&5.69 \\
H$_{C2}^{P}$(0) & T & 8.28& 10.34 &11.59 &12.32&12.14 \\
H$_{C2}^{orb}$(0) & T &3.34 & 5.89&6.77&6.30&3.68 \\
$\alpha_{m}$ &  & 0.56& 0.80 &0.82 &0.72&0.43 \\
$\xi_{GL}$& $nm$ & 6.52 & 5.62&5.95&5.96&7.60 \\
$\lambda_{GL}^{mag}$& $nm$&150.74 & 144.64&151.93&156.84&149.19\\
$k_{GL}$& &23.07 & 25.73&25.53&26.31 &19.63\\
$\rho_{300K}/\rho_{8K}$(RRR) & & 1.48 & 1.02&1.11&1.20&1.59 \\
$\gamma_{n}$& mJ mol$^{-1}$ K$^{-2}$& 4.44 & 5.87 & 7.76 & 3.51 & 3.32 \\
$\theta_{D}$& K& 164.66 &158.73&146.50&170.92&190.20\\
$\frac{\Delta_{sp}(0)}{k_{B}T_{C}}$ & & 1.87 & 2.56&2.45&2.61&2.14 \\
$\lambda_{e-ph}$ &  &0.75&0.84&0.91&0.88&0.82\\
$\frac{m*}{m_{e}}$ &  & 0.7&0.8 &0.94&0.6&0.58 \\
$v_{f}$  & 10$^{5}$ ms$^{-1}$ &7.54&7.61 &7.19& 8.11&8.15 \\
$\frac{\xi_{0}}{l_{e}}$ &  & 2.7&3.58 & 4.87& 2.66&2.15\\
$\frac{T_{C}}{T_{F}}$  & &0.00034&0.00036&0.00038& 0.00051&0.00051\\
\hline
\end{tabular}
\par\medskip\footnotesize
\end{center}
\end{table*}

Using the known value of $\beta$, the Debye temperature $\theta_{D}$ can be calculated using the expression $\theta_{D} = \left(\frac{12\pi^{4} R n}{5\beta}\right)^{\frac{1}{3}}$, where $R$ is the universal gas constant (8.314 J K$^{-1}$ mol$^{-1}$) and $n = 1$ is the number of atoms per unit cell. The density of states (DOS) at the Fermi level, denoted as $D(E_F)$, can be determined using the relation $D(E_F) = \frac{3\gamma_n}{\pi^{2} k_{B}^{2}}$, where $k_{B}$ is the Boltzmann constant \cite{Book}. By applying these equations, the calculated values of $\theta_{D}$ are lower than the Debye temperature of the corresponding elements. Furthermore, as the concentration increases up to 50\% of the Ta content, the values of $\theta_{D}$ decrease, indicating that doping introduces some disorder in binary alloys. The values obtained from $\theta_{D}$ are provided in \tableref{tbl: SC parameters}.
Subsequently, we determined the electron-phonon coupling constant, denoted as $\lambda_{e-ph}$. This dimensionless constant quantifies the attractive interaction between electrons and phonons and can be calculated using the values of $T_{C}$ and $\theta_{D}$ through the semi-empirical McMillan formula \cite{Mcmillan},

\begin{equation}
\lambda_{e-ph} = \frac{1.04+\mu^{*}\mathrm{ln}(\theta_{D}/1.45T_{C})}{(1-0.62\mu^{*})\mathrm{ln}(\theta_{D}/1.45T_{C})-1.04}.
\label{eqn7:Lambda}
\end{equation} 
where the repulsive screened Coulomb parameter, denoted as $\mu^{*}$, typically falls within the range of 0.09 to 0.18, with a commonly used value of 0.13 for intermetallic compounds. In our case, we adopted a value of 0.13. The estimated values of $\lambda_{e-ph}$ for Ta$_{x}$Hf$_{1-x}$ alloys range from 0.7 to 1, as summarized in \tableref{tbl: SC parameters}. Similar values of $\lambda_{e-ph}$ have been observed in other binary compounds, such as Re$_{1-x}$Mo$_{x}$ \cite{Re-Mo}. The higher values of $\lambda_{e-ph}$ in Ta$_{x}$Hf$_{1-x}$ alloys indicate a strong coupling strength among electrons in the superconducting state \cite{Ta-Nb}.

The inset of \figref{Fig4.jpg} presents the electronic component of the specific heat, denoted as C$_{el}$ (shown for $x$ = 0.2 and 0.5), which was obtained by subtracting the contribution of the phonon from the experimental data using the relation C$_{el}$(T) = C - $\beta $T$^{3}$. The C$_{el}$(T) data for all compositions are well described by the isotropic s-wave model in the BCS theory, as outlined in ref. \cite{BCS2}. The black line in the inset of \figref{Fig4.jpg} represents the curve fitted to the data using the single-gap s-wave BCS equation. The values obtained from the superconducting gap ($\frac{\Delta(0)}{k{B}T_{C}}$), listed in \tableref{tbl: SC parameters}, exceed the predicted BCS value, indicating the presence of strongly coupled superconductivity in Ta$_{x}$Hf$_{1-x}$ compounds.

\subsection{Electronic properties and the Uemura classification}
The Sommerfeld coefficient $\gamma_{n}$, is related to the effective mass of quasi-particles $m^{*}$ and electronic carrier density $n$ of the system by the following expression \cite{Book},
\begin{equation}
\gamma_{n} = \left(\frac{\pi}{3}\right)^{2/3}\frac{k_{B}^{2}m^{*}V_{\mathrm{f.u.}}n^{1/3}}{\hbar^{2}N_{A}}
\label{eqn8:gf}
\end{equation}

where k$_{B}$ = 1.38 $\times$10$^{-23}$ J/K is the Boltzmann's constant, N$_{A}$ and V$_{\mathrm{f.u.}}$ are the Avogadro number and the volume of a formula unit, respectively. 
The following relations can be used to connect the electronic mean free path $l_{e}$ and the carrier density $n$ to the Fermi velocity $v_{\mathrm{F}}$ and the effective mass m$^{*}$,

\begin{equation}
\textit{l}_{e} = \frac{3\pi^{2}{\hbar}^{3}}{e^{2}\rho_{0}m^{*2}v_{\mathrm{F}}^{2}}, \quad n = \frac{1}{3\pi^{2}}\left(\frac{m^{*}v_{\mathrm{F}}}{\hbar}\right)^{3}
\label{eqn9:le,n}
\end{equation}

In the dirty limit, the GL penetration depth $\lambda_{GL}$(0) and coherence length $\xi_{GL}$(0) get affected, which can be described in terms of London penetration depth ($\lambda_{L}$) and BCS coherence length ($\xi_{0}$), by the following modified Eqs.\ref{eqn10:f} and Eqs.\ref{eqn11:xil}, respectively,
\begin{equation}
\lambda_{GL}(0) = 
\lambda_{L}
\left(1+\frac{\xi_{0}}{\textit{l}_{e}}\right)^{1/2}, \quad \lambda_{L} =
\left(\frac{m^{*}}{\mu_{0}n e^{2}}\right)^{1/2}
\label{eqn10:f}
\end{equation}
\begin{equation}
\frac{\xi_{GL}(0)}{\xi_{0}} = \frac{\pi}{2\sqrt{3}}\left(1+\frac{\xi_{0}}{\textit{l}_{e}}\right)^{-1/2}
\label{eqn11:xil}
\end{equation}

\begin{figure}
\includegraphics[width=1.0\columnwidth]{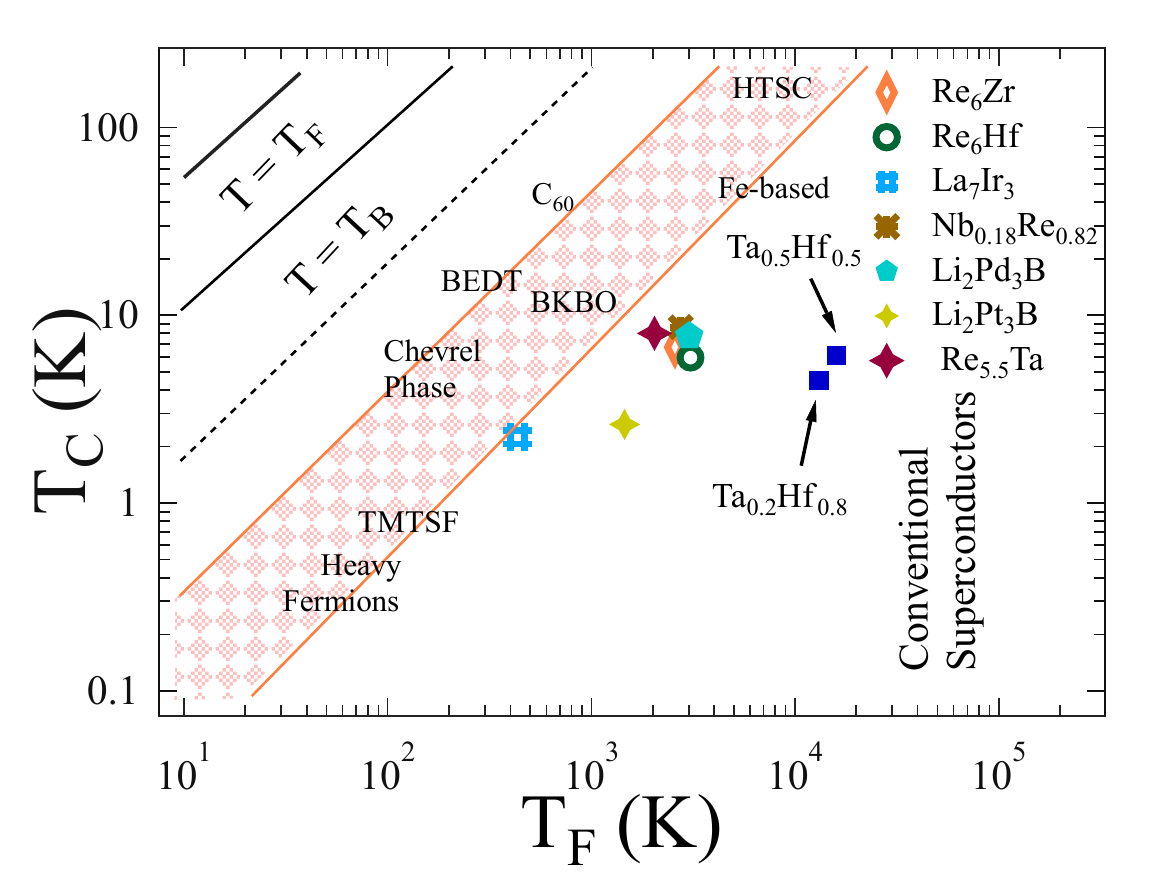}
\caption{\label{Fig5.jpg} An "Uemura plot" showing superconducting transition temperature (T$_{C}$) versus effective Fermi temperature (T$_{F}$), in which blue square symbols representing the positions of Ta$_{x}$Hf$_{1-x}$. The data point is inside two solid boundary lines depicting the unconventionality zone.} 
\end{figure}

The above set of Eqs.(\ref{eqn8:gf}-\ref{eqn11:xil}) were solved simultaneously as done in Ref. \cite{Umera_Ref} to evaluate the electronic parameters $n$, m$^{*}$, $\xi_{0}$ and $l_{e}$ using the values of $\gamma_{n}$, $\xi_{GL}$(0), and $\rho_{0}$ for varying the chemical compositions of Ta$_{x}$Hf$_{1-x}$. Finally, the effective Fermi temperature (T$_{F}$) for Ta$_{x}$Hf$_{1-x}$ was calculated using the expression \cite{Tf}, where $n$ and $m^{*}$ are the electronic carrier density and the effective mass of quasi-particles, respectively:
\begin{equation}
 k_{B}T_{F} = \frac{\hbar^{2}}{2}(3\pi^{2})^{2/3}\frac{n^{2/3}}{m^{*}}, 
\label{eqn12:tf}
\end{equation}

We have compiled all the estimated electronic parameters of Ta$_{x}$Hf$_{1-x}$ in \tableref{tbl: SC parameters}. The ratio $\xi_{0}$/$l_{e}$ exceeds the expected range for clean limit superconductivity, indicating that the Ta-Hf binary alloy exhibits dirty limit superconductivity.

Uemura et al. \cite{Uemura} have established a distinction between conventional and unconventional superconductors based on the $\frac{T_{C}}{T_{F}}$ ratio. When this ratio falls within the range of 0.01 $\leq$ $\frac{T_{C}}{T_{F}}$ $\leq$ 0.1, the material is classified as an unconventional superconductor. Unconventional superconductors encompass various superconducting compounds, including heavy fermionic materials, Chevrel phases, high T$_{C}$ superconductors, iron-based superconductors, and other exotic superconductors. The boundary for unconventional superconductors is represented by two solid peach lines in \figref{Fig5.jpg}, and the values of $\frac{T_{C}}{T_{F}}$ for each composition (indicated by the blue symbols) lie between 0.00034 and 0.00052. These values position the Ta$_{x}$Hf$_{1-x}$ compounds outside the unconventional superconductivity region.

\begin{figure}
\includegraphics[width=1.0\columnwidth]{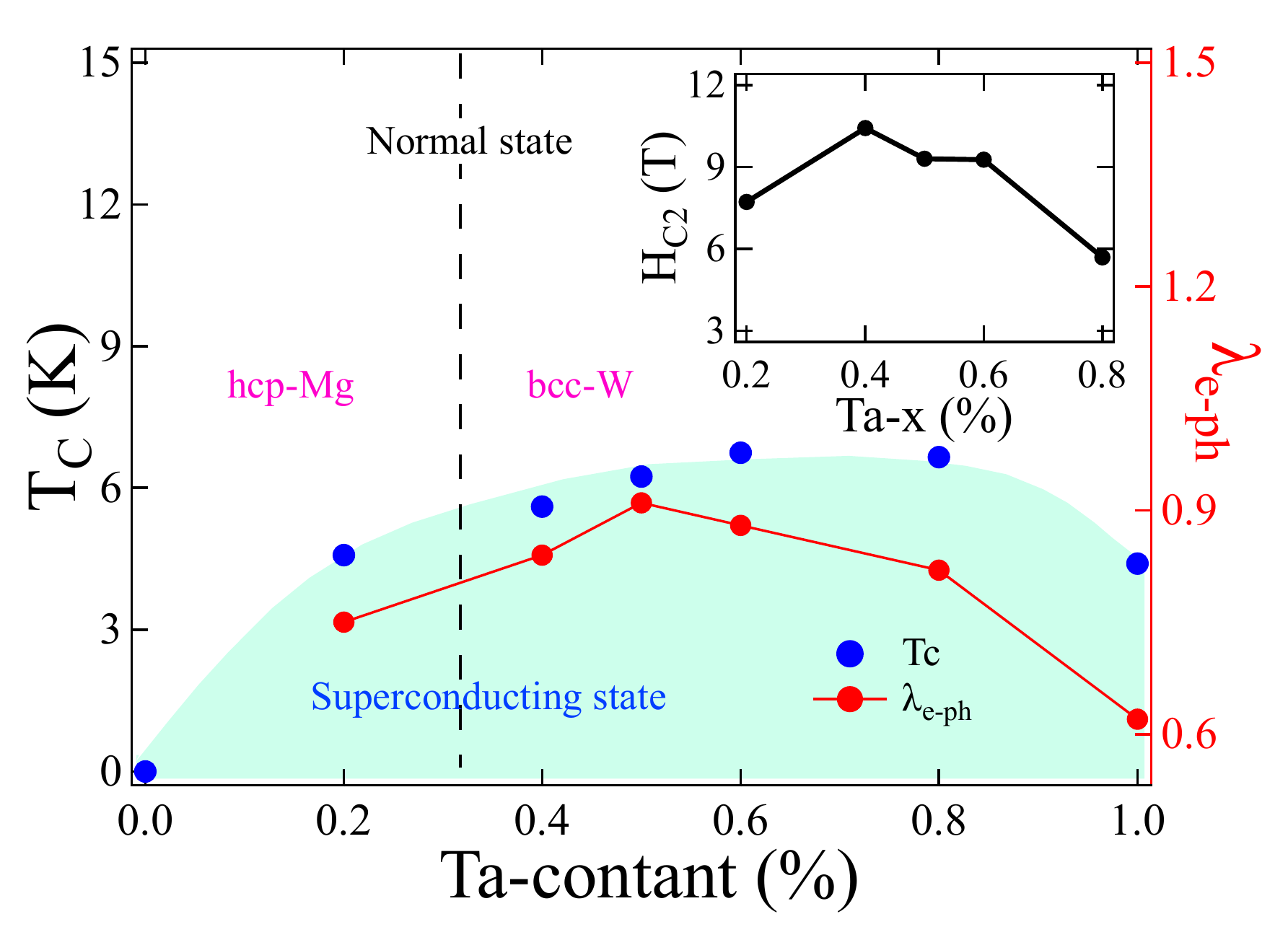}
\caption {\label{Fig6.phase diagram} Superconducting phase diagram of the Ta$_{x}$Hf$_{1-x}$ alloys. Superconducting transition temperatures T$_{C}$ and electron-phonon coupling $\lambda_{e-ph}$ vs Ta content. Inset shows a variation of an upper critical field with changing Ta content.}
\label{Phase diagram.jpg}
\end{figure}

\section{PHASE DIAGRAM}
The experimental data obtained from the superconducting Ta$_{x}$Hf$_{1-x}$ binary alloys are summarized in the phase diagram shown in \figref{Fig6.phase diagram}. As the relative content of Ta and Hf is varied, two crystal structures are observed: W-$bcc$ (I$m\bar{3}m$) and Mg-$hcp$ (P6$_{3}$/$mmc$). Crystal structure transitions occur between $x$ = 0.2 and 0.4 at the phase boundary. The highest and lowest values of T$_{C}$, 6.7 K and 4.5 K, respectively, are obtained for samples with $x$ = 0.6 (W cubic-type structure) and $x$ = 0.2 (Mg-$hcp$ type structure), as depicted in \figref{Fig6.phase diagram}. The variation of $x$ on both sides leads to a decrease in the transition temperature, which can be attributed to the influence of elemental Ta and Hf on their respective sides \cite{Ta, Ta-Hf}.

The phase diagram also shows the variation of the electron-phonon coupling constant with the Ta content, exhibiting a dome-like behavior. The three series of BCC alloys (3d, 4d, 5d) demonstrate the same dome behavior, with a peak at a valence electron/atom ratio $n$ = 4.5, a deep minimum near $n$ = 5.8, and a shoulder at $n$ = 6.2 \cite{Mcmillan}. The enhanced values of the electron-phonon coupling constant in Ta-Hf binary alloys are also higher than those of the elemental Hf and Ta, indicating strongly coupled superconductivity. The inset of \figref{Fig6.phase diagram} illustrates the variation between the upper critical field H$_{C2}(0)$ and the Ta/Hf concentration. A similar trend has been observed in previous studies of other binary solid solution alloys \cite{HC2_TM}. Ta$_{0.4}$Hf$_{0.6}$ corresponds to the highest value of $H_{C2}(0)$, which is 10.43 T (comparable to the Pauli paramagnetic limit), and as the concentration of Ta increases, H$_{C2}(0)$ decreases in the phase diagram. 

\section{Conclusion}
This study presents experimental results focusing on Ta$_{x}$Hf$_{1-x}$ ($x$ = 0.2, 0.4, 0.5, 0.6, 0.8) binary alloys. Through powder X-ray diffraction (XRD) analysis, it was determined that the crystal structures of Ta$_{x}$Hf$_{1-x}$ alloys encompass two phases across the entire range of solid solution: W-$bcc$ (for $x \geq 0.4$) and Mg-$hcp$ (for $x < 0.4$). Investigation of magnetization, electrical transport, and thermodynamic properties revealed that Ta$_{x}$Hf$_{1-x}$ alloys exhibit superconductivity, with the highest bulk transition temperature observed at T$_{C}$ = 6.7 K for the composition $x$ = 0.6. Significantly higher calculated values of the upper critical field H$_{C2}(0)$ were obtained, particularly for $x$ = 0.2 and 0.4, indicating their proximity to the BCS Pauli limiting field. Furthermore, analysis of the low-temperature specific heat data indicated a fully gapped superconducting state, with a larger gap magnitude exceeding the BCS value of 1.76. This magnitude is comparable to unconventional superconductors based on 5d compounds, suggesting an enhanced electron-phonon coupling constant in Ta$_{x}$Hf$_{1-x}$ alloys. The unique characteristics of Ta$_{x}$Hf$_{1-x}$ alloys, such as their higher upper critical fields, the presence of a fishtail characteristic (indicating unusual lattice vortices), larger superconducting gap values and strong electron-phonon coupling, suggest possible unconventional behavoiur and make them intriguing for potential applications in superconducting devices. However, further investigations are necessary using microscopic probes such as nuclear magnetic resonance (NMR) and muon spin resonance ($\mu$SR) on single/polycrystalline samples. These investigations will provide insights into the superconducting pairing mechanism and a better understanding of the possible unconventional nature and anomalous enhancement of the superconducting transition in the Ta$_{x}$Hf$_{1-x}$ binary alloy.


\section{Acknowledgments} 
P. K. Meena acknowledges the funding agency Council of Scientific and Industrial Research (CSIR), Government of India, for providing the SRF fellowship (Award No: 09/1020(0174)/2019-EMR-I). R. P. S. acknowledges the Science and Engineering Research Board, Government of India, for the Core Research Grant CRG/2019/001028.

\end{document}